\documentclass[aps,prb,reprint,nofootinbib,superscriptaddress,floatfix]{revtex4-2}
\usepackage{graphicx} 
\usepackage{braket}
\usepackage{xcolor}
\usepackage{amsmath}
\usepackage{makecell}
\usepackage{float} 
\usepackage{graphicx}
\usepackage{dcolumn}
\usepackage{bm}
\usepackage{braket}
\usepackage{hyperref}
\usepackage{amssymb}
\usepackage{soul}
\usepackage{comment}

\begin{document}

\title{
Floquet thermalization by power-law induced permutation symmetry breaking}

\author{Manju C}
\email{222004003@smail.iitpkd.ac.in}

\author{Uma Divakaran}
 \email{uma@iitpkd.ac.in}
\affiliation{
Department of Physics, Indian Institute of Technology Palakkad, Palakkad, Kerala-678623.}

\date{\today}
\begin{abstract}
Permutation symmetry plays a central role in the understanding of collective quantum dynamics. By introducing power law couplings that algebraically decay with the distance between the spins $r$ as $1/r^{\alpha}$, we break this symmetry with a non-zero $\alpha$. \textcolor{black}{This allows us to probe the emergence of new dynamical behaviors, including thermalization in an otherwise permutation symmetric Hamiltonian with all-to-all spin interactions along $x$ direction subjected to periodic kicks in transverse direction.} 
 As we increase $\alpha$, the system interpolates from an infinite range spin system at $\alpha=0$ exhibiting permutation symmetry, to a short range integrable model as $\alpha \rightarrow \infty$ where this permutation symmetry is absent. We focus on this change in the behavior of the system as $\alpha$ is tuned, using dynamical quantities like total angular momentum and von Neumann entropy. Starting from the chaotic limit of the permutation symmetric Hamiltonian at $\alpha=0$, \textcolor{black}{for the finite system sizes considered,} we find that for small $\alpha$, the steady state values of these quantities remain close to the permutation symmetric subspace values corresponding to $\alpha=0$. At intermediate $\alpha$ values, these show signatures of thermalization exhibiting  values corresponding to that of random states in full Hilbert space. On the other hand, the large $\alpha$ limit approaches the values corresponding to integrable kicked Ising model. In addition, we also study the dependence of thermalization on the driving period $\tau$, with results indicating the onset of thermalization for smaller values of $\alpha$ when $\tau$ is large, thereby extending the thermalizing window in the intermediate range of $\alpha$. We further confirm these results using effective dimension 
and spectral statistics.

\end{abstract}

\maketitle

\section{Introduction}

Long range interacting spin systems have known to exhibit many intriguing properties such as instantaneous spreading of correlations  \cite{prl_long_range}, presence of thermalization plateaus \cite{Worm_2013}, ergodicity breaking and long relaxation times \cite{ergodicity_breaking}. 
Periodically driven systems have garnered more interest recently owing to its existence of non trivial Floquet phases that are not present in equilibrium systems. The ability to engineer material properties using controlled pulses have also made the periodic drive experimentally feasible as in the case of trapped ions, ultracold atoms, cavity QED systems etc \cite{trapped_ion_exp, prl_ultra_cold, cavity_QED}. In the case of periodically driven many body system, where there is no conservation of energy, the system is expected to continuously heat ultimately leading to an infinite temperature state \cite{infinite_temp_PRX, anushya_infinite_temp, PRE_RMT_infinite_temp}. However, there are studies which show how this heating can be controlled by tuning the driving frequency \cite{exponential_slow_heating}. 
In many cases, it is seen that when the driving frequency is large compared to the local energy scales in the system, thermalization to infinite temperature state takes longer, resulting to prethermal phases. The lifetime of these prethermal phases depends exponentially on the frequency of the drive. The dynamics of various observables in this prethermal phase can be determined by a local Floquet Hamiltonian \cite{Prethermal_conserved_PRB, christoph_prethermal}. Experimentally, Floquet thermalization have been observed using NMR techniques in a nuclear spin system \cite{exp_prethermal} and in a bosonic cloud of ultracold atoms \cite{expt_bosonic_prethermal}. There are also cases, where the heating is suppressed by introducing disorder that leads to many body localization in periodically driven systems \cite{prl_MBL, alessio}.
On the other hand, when the driving frequency is small, the absorption from the drive is possible resulting in thermalization to an infinite temperature state described by random matrix theory \cite{thermalisation_rmt, heating_timescale_souvik, PRE_RMT_infinite_temp}.

We now integrate the study of periodic driving with the field of long range interacting systems where the range of interaction is tunable. 
\textcolor{black}{In particular, the system is described by a Hamiltonian consisting of spin-spin interactions along $x-$ direction with the interaction strength decaying as $1/r^{\alpha}$, where $r$ is the distance between the two spins, and $\alpha$ is a positive constant. In addition, the system is subjected to periodic kicks in the transverse direction.} In the limit of all-to-all spin interactions corresponding to $\alpha=0$, where there exists a permutation symmetry in the system, the dynamics can be studied using collective degrees of freedom \cite{tripartitescrambling, chinni_pspin_models, olsacher2022digital}. As the range of interactions is decreased or $\alpha$ is increased, this symmetry is broken and it is this symmetry breaking that we focus in the current work. For large $\alpha$, only the nearest neighbour interactions are dominant leading to the well studied kicked short range interacting system \cite{ising_particle_duality, kicked_ising_arul, kicked_ising_maxent, banaras_kicked_ising}. Thus, the system interpolates between long range and short range as the strength of the interaction is tuned. These kind of power law interactions are naturally present in spin glasses and magnetically frustrated systems \cite{spin_glass_powerlaw, magnetic_frustrated}. They can also be engineered in the case of atomic, molecular and optical systems  such as trapped ions, dipolar systems and quantum gases in cavities \cite{yan2013observation, trapped_ion_prl, Rev_mod_Rydberg_atoms, britton2012engineered, jurcevic2014quasiparticle}.  Several studies have been carried out on the undriven power law interacting spin systems focusing entanglement dynamics \cite{Pappalardi_scrambling, variable_prx}, information propagation \cite{light_cone_pra, correlations_exp}, phase transitions \cite{prl_phase_transitions}, thermalization \cite{exp_power_law, liu2019confined, ensemble_inequivalence, undriven_pizzi, pal2025fragmented}, quantum many-body scars \cite{many_body_scars} etc. These studies suggest that the \textcolor{black}{undriven} long range spin chain in one dimension can be divided into three distinct regimes: (\romannumeral 1) $\alpha<1$, the near infinite range regime, (\romannumeral 2) $1<\alpha<2$ the intermediate regime where the interactions are still long range and (\romannumeral 3) $\alpha>2$, the short range interaction regime \cite{prl_long_range, Pappalardi_scrambling, aditi_sen, long_range_prethermal_phases}.
Previous studies on periodically driven long range interacting spin chains have focused on thermalization time scales, stability of dynamical phases and suppression of heating \cite{kapitza, santos_heating_suppression, exponential_slow_heating}.
In contrast, the main focus of this study is to investigate the effect of breaking permutation symmetry in the original all-to-all interacting system by introducing power law interactions that decay as $1/r^{\alpha}$ in a kicked spin system. Specifically, we study how increasing $\alpha$ causes the dynamics to leave the symmetric subspace at $\alpha=0$ and extend beyond it, using both dynamical quantities like the total angular momentum operator $J^2$ and von Neumann entropy $S_{N/2}$ of subsystem of size $N/2$, and those related to Floquet eigenstates. 
\textcolor{black}{We also comment on different dynamical regimes that emerge in the presence of periodic drive for a finite sized system based on the thermalization behavior of the system.}
We performed a similar study when the permutation symmetry is broken by introducing disorder in an all-to-all interacting kicked spin chain of $N$ qubits. The results showed the presence of a continuous phase transition from a phase where the dynamics is restricted within the permutation symmetric subspace (PSS) of $N+1$ dimensions to the phase where the dynamics explored the entire $2^N$ Hilbert space dimension as the strength of disorder is increased \cite{manju_2025_SR}. We also observed the convergence to random matrix theory corresponding to the full Hilbert space (FHS) in the limit of large disorder \cite{manju_2025_SR}. 

The paper is organised as follows: In Sec~\ref{model}., we introduce the long range interacting  Hamiltonian with periodic kicks in the transverse direction. Sec~\ref{dynamics}. consists of numerical results \textcolor{black}{of dynamical quantities like $J^2$ and  $S_{N/2}$ for a fixed system size $N$ as a function of $\alpha$ as well as driving period $\tau$} . We also try to provide an analytic framework to understand these quantities in the case of large $\alpha$, i.e, kicked Ising model. Sec.~\ref{floquet_eigenstates} deals with eigenstate properties characterized by $D_{\rm eff}$ (defined later) and spectral statistics to show the transition from Wigner Dyson to Poisson as the interaction range $\alpha$ is increased. \textcolor{black}{In Sec \ref{sec_N}, we study the system size dependence  of various quantities}. Finally we conclude in Sec~\ref{conclusion}.

\section{Model}
\label{model}
The long range Hamiltonian consisting of $N$ spins  is given by:
\begin{equation}
H=\frac{k}{4N(\alpha) }\sum_{i\neq j}^{N}\frac{\sigma_{i}^{x}\sigma_{j}^{x}}{D_{ij}^\alpha} +\frac{p}{2}\sum_{n=-\infty }^{\infty }\sum_{i=1}^{N}\sigma_{i}^{z}\delta \left (\frac{t}{\tau}-n \right),
\label{eq_spin}
\end{equation}
where $N(\alpha)=\frac{1}{N-1}\sum_{i,j, i\neq j}\frac{1}{D_{i,j}^{\alpha}}$ is the rescaling factor also called the Kac factor, which preserves the extensivity of the Hamiltonian.   Here, $\sigma_i^{x},~\sigma_i^z$ are the Pauli matrices at site $i$, interaction strength for the long range coupling is denoted by $k$, $p$ represents the strength of the kicks in $z$ direction and $\tau$ is the time period between the kicks. We consider open boundary conditions by setting $D_{i, j}=|i-j|$. The parameter $\alpha$ controls the range of interaction where the two well studied extreme limits correspond to the
(i) kicked top model when $\alpha=0$ and (ii) kicked Ising model when $\alpha \to \infty$.

The kicked top model is a widely studied system in the field of quantum chaos \cite{haake1987classical, Haakebook}. It has been studied both, as a single top characterized by total angular momentum $J$, and as a generic all-to-all interacting spin system described by a single collective spin operator given by $J_{x}=\sum_{i=1}^{N}\sigma_{i}^{x}/2$, with similar expressions for $J_y$ and $J_z$ \cite{concurrence}. In the classical limit $(N\rightarrow \infty)$, tuning the interaction parameter $k$ of the Hamiltonian takes it from regular $(k<2)$ to chaotic dynamics $(k>4)$ \cite{decoherence}. It is well known that when the system is initialized in a spin coherent state with \textcolor{black}{$j=N/2$, where $j$ is the angular momentum quantum number}, the dynamics under the evolution of kicked top Hamiltonian gets restricted within the $N+1$ dimensional Hilbert space. This is due to conservation of $J^2(=J_x^2+J_y^2 + J_z^2)$ that follows from the permutation symmetry of the underlying system. This model is extensively studied in the context of entanglement dynamics \cite{concurrence, periodicity, spinsqueezing}, quantum-classical correspondence \cite{periodic_orbits}, and information scrambling \cite{tripartitescrambling}. There have been studies employing random matrix theory (RMT) to understand the chaotic limit ($k > 4$) of kicked top model \cite{tripartitescrambling}.\\

The other limit of $\alpha \rightarrow \infty$ corresponds to the kicked Ising model which is an integrable model \cite{Prosen_integrable_ising, PRE_ergodicity_ising, kicked_ising_arul, kicked_ising_maxent}. In case of periodic boundary conditions, it can be reduced to fermionic model and can be diagonalised using Jordan Wigner transformation. It is shown that the expectation values of observables in the long time limit show convergence to periodic Generalised Gibbs ensemble (GGE) \cite{banaras_kicked_ising}. In the kicked Ising case, the effect of integrability breaking, such as addition of a longitudinal field that results in transition from integrable to non integrable systems, has also been examined in previous studies \cite{kicked_ising_arul, sff_ising, ising_particle_duality, Rohit_sunil_mishra_ising}. 

In this work, we focus on the effect of increasing $\alpha$ from long range limit ($\alpha=0$) to short range limit (large $\alpha$) on various dynamical quantities. The evolution operator for the Hamiltonian in Eq.~\ref{eq_spin} is given by:
\begin{equation}
U=\exp{\left (-\frac{ik\tau}{4N(\alpha)} \sum_{i\neq j}^{N}\frac{\sigma_{i}^{x}\sigma_{j}^{x}}{D_{ij}^\alpha}\right)} \exp{\left(-i\frac{p \tau}{2}\sum_{i=1}^{N}\sigma_{i}^{z}\right)}.
\label{eq:Floquet}
\end{equation}

While $\alpha=0$ corresponds to fully permutation symmetric model, any non zero $\alpha$ breaks this symmetry. We now examine the effect of this permutation symmetry breaking by a non-zero $\alpha$ using quantities like total angular momentum $J^2$ and entanglement entropy $S_{N/2}$. \textcolor{black}{We set $p=4\pi/11$ for which the Hamiltonian has reduced number of symmetries making it convenient for the spectral analysis done later. All the calculations are performed with the chaos parameter $k=6$ corresponding to the chaotic regime of $\alpha=0$ limit, where RMT results within the permutation-symmetric subspace have been  studied extensively in the literature \cite{tripartitescrambling}.} We also consider different values of driving period, namely $\tau=0.1,1.0$ and $5.0$.
We start from an initial spin coherent state defined as \cite{Haakebook, spincoherentstates, glauber_spin_coherent_state}: 
\begin{equation}
 \ket{\psi_0}=|\theta, \phi\rangle= \left(\cos\frac{\theta}{2} \ket{0}+e^{i\phi}\sin \frac{\theta}{2} \ket{1}\right)^{\otimes N}, 
 \end{equation}
 with fixed values of $\theta=2.25,\phi=1.1$. The state of the system after $n$ time steps is given by $\ket{\psi_n}=U^n\ket{\psi_0}$. In order to obtain the steady state values, the time average is taken from $10^5$ to $3 \times 10^5$ unless specified explicitly. We 
 discuss the evolution of $J^2$ and $S_{N/2}$ \textcolor{black}{for system size $N=14$ in the next section, and later discuss its dependence on $N$}.
 
\section{Dynamical Quantities}
\label{dynamics}
\subsection{Total angular momentum $J^2$}
\begin{figure}[ht]
\centering
     \includegraphics[width=1.0\linewidth, height=0.8\linewidth]{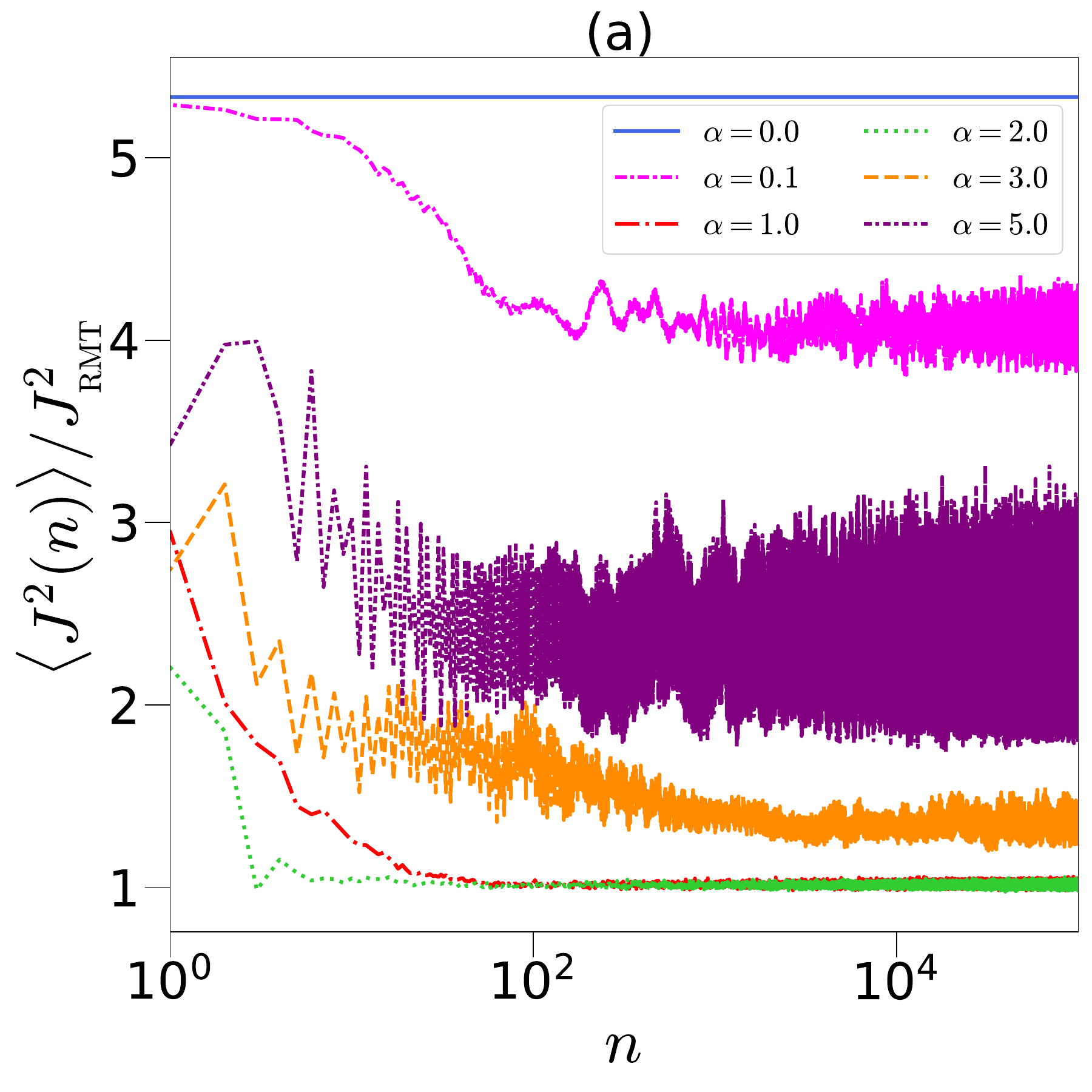}
   \includegraphics[width=1.0\linewidth, height=0.8\linewidth]{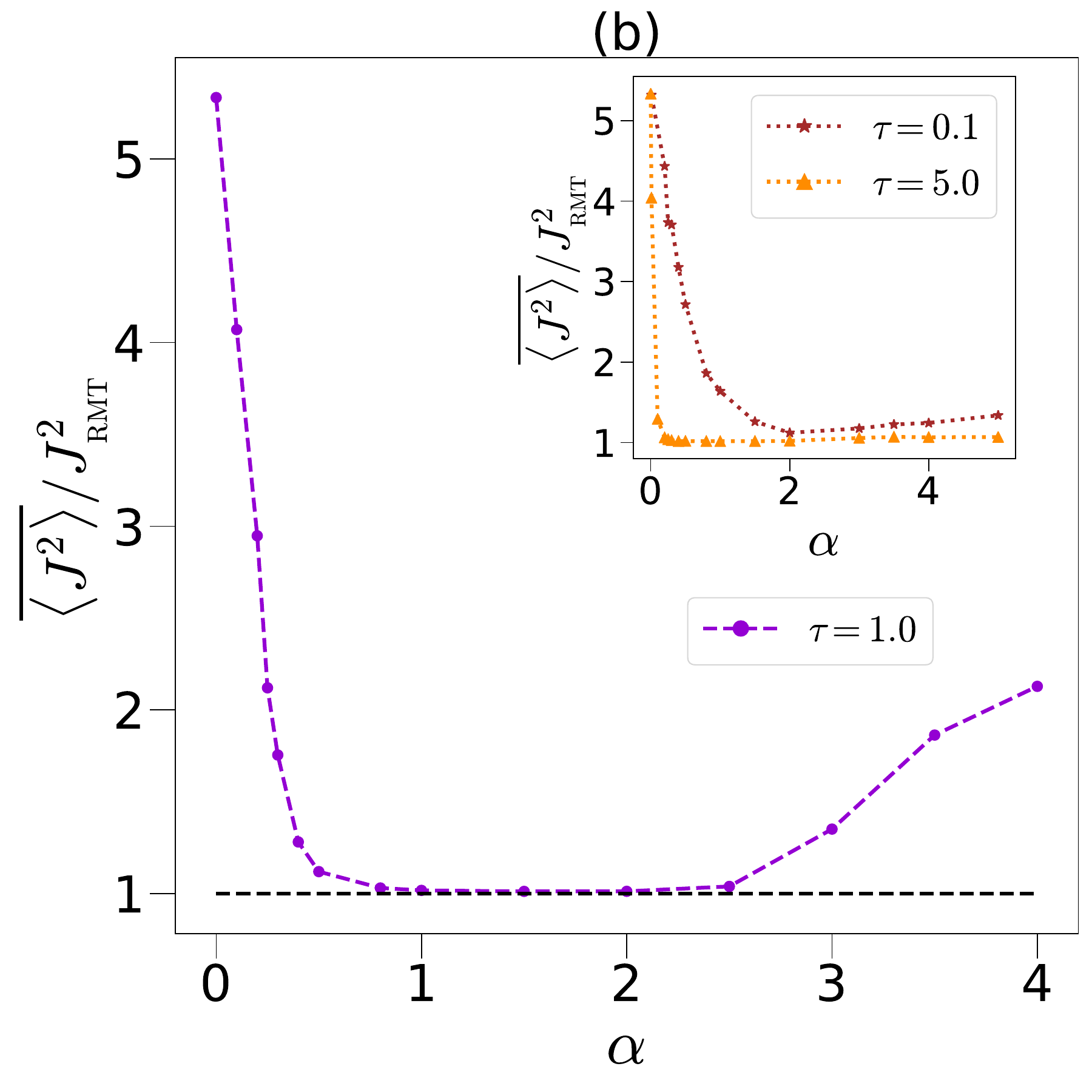}
   \caption{(a): \textcolor{black}{The time evolution of $\left \langle J^2 \right \rangle$ rescaled by $J^2 _{{RMT}}$ for $N=14$ at $\tau=1.0$ for different $\alpha$ values. (b): The time averaged $J^2$ rescaled by $J^2_{\rm_{RMT}}$ as a function of $\alpha$ for the same $\tau$ and $N$. The horizontal black dashed line correspond to $\langle J^2 \rangle/J^2_{RMT}=1$, clearly indicating saturation to RMT values ($=3N/4$) for the intermediate $\alpha$ range. The inset shows the plot of $\langle J^2 \rangle/ J^2_{RMT}$ with respect to $\alpha$ for $\tau=0.1$ and $\tau=5.0$ showing thermalization over a wider intermediate range of $\alpha$ for larger $\tau$.}}

  \label{J2_timeavg}
\end{figure}

The evolution of total angular momentum operator $J^2$ at any time $n$, defined as $\langle J^2(n) \rangle=\langle {\psi_n|J^2|\psi_n \rangle}$, \textcolor{black}{scaled by the corresponding RMT value $J^2_{\rm_{RMT}}$ discussed later}, is shown in Fig.~\ref{J2_timeavg}(a) for different $\alpha$, where the period of driving $\tau$ is set to unity. As stated earlier, $\alpha=0$ is the kicked top model where $\langle J^2 \rangle$ is a constant of motion and its value is given by $\langle J^2 \rangle=N/2(N/2+1)$. For any non-zero $\alpha$, $J^2$ is no longer a constant of motion and its value decreases from that at $\alpha=0$, and reaches a steady state at large $n$. The time averaged steady state value denoted by $\overline{\langle J^2 \rangle}$, \textcolor{black}{after rescaling}, is plotted as a function of $\alpha$   in Fig.~\ref{J2_timeavg}(b).
We see that for $\alpha$ small, $\overline{\langle J^2 \rangle}$ stays close to $N/2 (N/2 +1)$ corresponding to the permutation symmetric subspace, and decreases as $\alpha$ is increased, reaching a minimum around $\alpha=0.5$. This minimum value extends till $\alpha=2$, after which the steady state value again increases. The minimum value attained in this case is equal to the one obtained for random states in full Hilbert space, given by $J^2_{\rm{RMT}}=3N/4$ \cite{manju_2025_SR},  see Fig.~\ref{J2_timeavg}(b). 
We also study the behavior of $\overline{\langle J^2 \rangle}/J^2_{\rm_{RMT}}$ for other values of $\tau$ which is shown in the inset of Fig.~\ref{J2_timeavg}(b). We observe a similar trend like the $\tau=1$ case, i.e, an initial monotonic decrease followed by an increase. As compared to $\tau=1$, we find that for larger $\tau$, the minimum value of $\langle J^2 \rangle$ is reached at smaller $\alpha$, and extends upto larger range of $\alpha$. For example, when $\tau=1$, the minima is reached at $\alpha=0.5$ which continues till $\alpha=2$, whereas for $\tau=5$, the minima is reached at $\alpha=0.1$ and extends till $\alpha=5$. This behavior is expected as decreasing the driving frequency (or increasing $\tau$) can make the system absorb energy from the drive, resulting in faster thermalization \cite{exponential_slow_heating}. The deviation of $J^2$ from $J^2_{\rm{RMT}}$ for large $\alpha$ is justified owing to its transition to a more localized short range interacting integrable spin chain at large $\alpha$ values. We shall explain this $\tau$ and $\alpha$ dependence using spectral statistics in sec \ref{sec_spectralstatistics}.

\subsection{von Neumann entropy  $S_{N/2}$}
\label{sec_vonNeumann}
\begin{figure}[ht]
\centering
   \includegraphics[width=1.0\linewidth, height=0.8\linewidth]{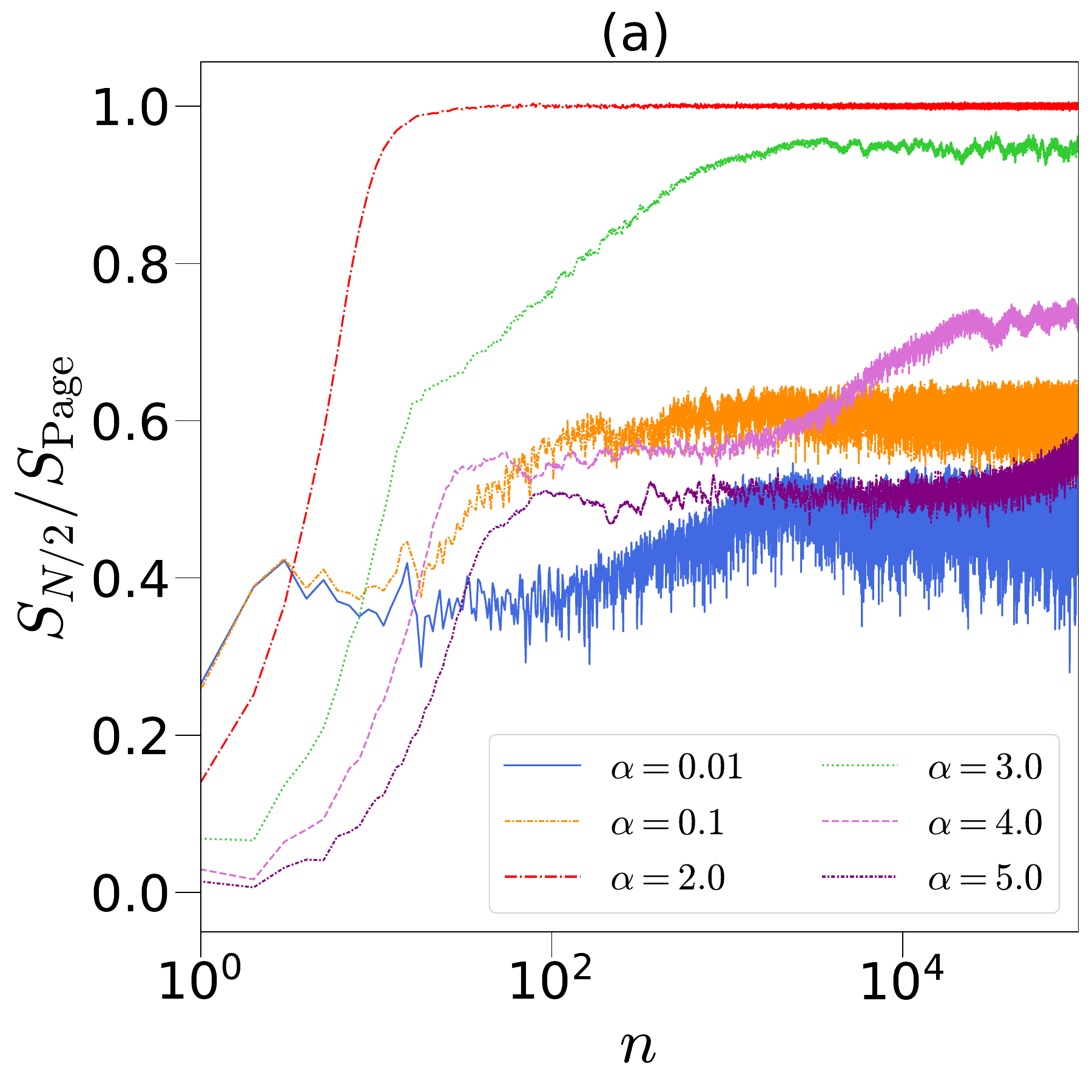}
   \includegraphics[width=1.0\linewidth, height=0.8\linewidth]{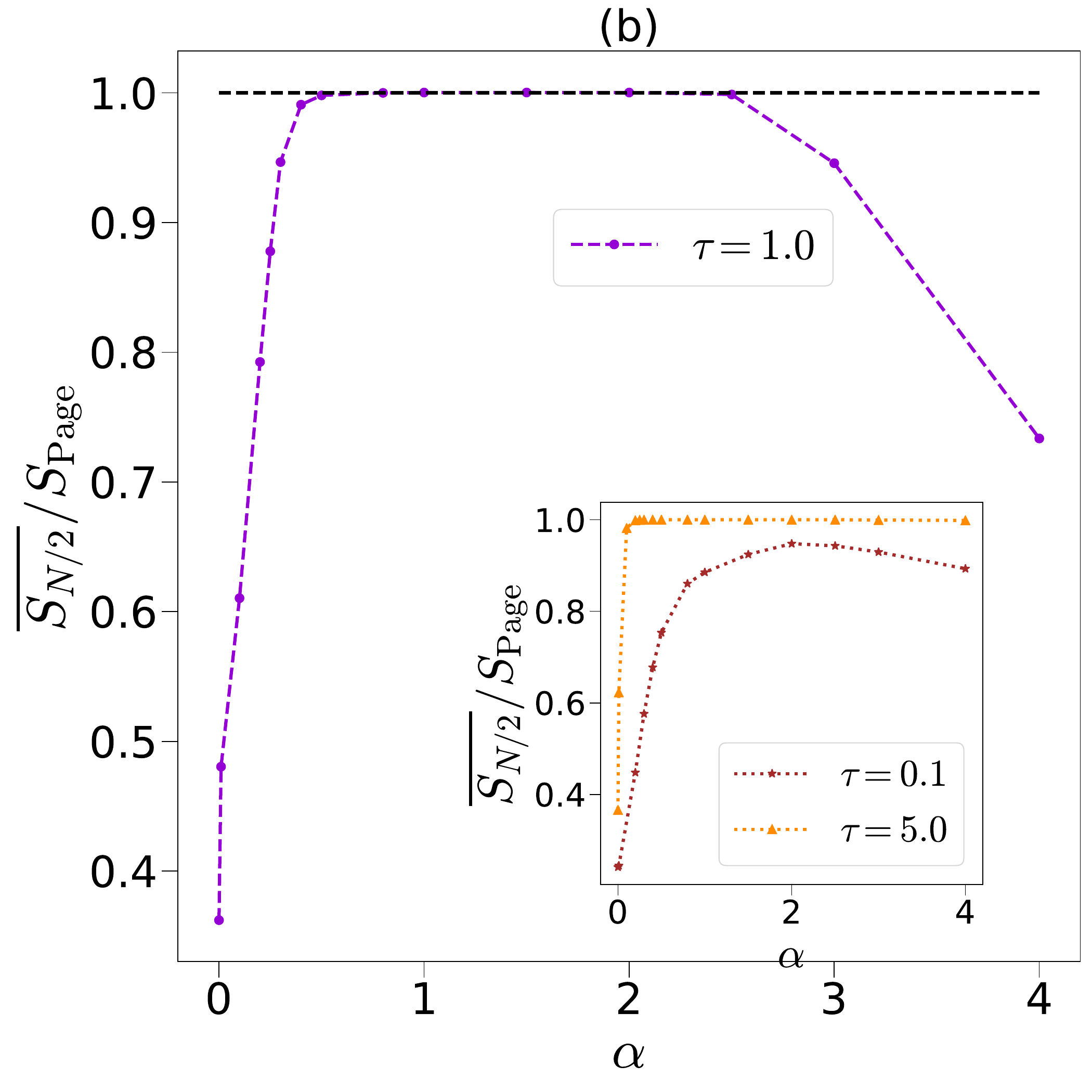}
   \caption{(a): \textcolor{black}{Time evolution of $S_{N/2}$  normalized with the Page value for $N=14$ at $\tau=1.0$ for different $\alpha$ values. (b): The plot of time averaged $S_{N/2}$ rescaled by $S_{\rm_{Page}}$ with respect to $\alpha$ for the same $\tau$ and $N$. The horizontal black dashed line corresponds to $\overline{S_{N/2}}/S_{\rm_{Page}}=1$, highlighting the saturation of $S_{N/2}$ to the Page value in the intermediate $\alpha$ range. The inset shows the same for $\tau=0.1$ and $\tau=5.0$, where we clearly see broadening of intermediate $\alpha$ range as $\tau$ is increased, see text for more details.}}

  \label{entropy}
\end{figure}

We further confirm the transition to RMT behavior at the intermediate range of $\alpha$ using von Neumann entropy.
The von Neumann entanglement entropy $S_{N/2}$ between a bipartition of $N/2:N/2$ qubits of an $N$ qubit system is given by:
\vspace{-5pt}
\begin{equation}
S_{N/2}=-\mathrm{tr}(\rho_{N/2} \log \rho_{N/2})
\end{equation}
where $\rho_{N/2}$ is the reduced density matrix of half-chain subsystem consisting of $N/2$ qubits. As stated earlier, we have fixed $k=6$, which corresponds to the chaotic region of the kicked top model. For $\alpha=0$, the entropy approaches the value predicted by RMT in the permutation symmetric subspace of $N+1$ dimensions, given by $\langle S_{N/2} \rangle_{PSS}\approx \log_2(N/2+1)- 2/3$ \cite{tripartitescrambling}. On the other hand, Page value given by $S_{\rm_{Page}}=\langle S_{N/2} \rangle_{FHS} ={N}/{2}-{1}/({2 \ln 2})$, corresponds to entropy for random pure states in full Hilbert space \cite{page1993average}. 
The plot of $S_{N/2}$ \textcolor{black}{scaled with the corresponding RMT value $S_{\rm_{Page}}$, as a function of $n$} for different $\alpha$ values is shown in Fig.~\ref{entropy}(a). For small $\alpha$ ($\alpha=0.01,~0.1$), it exhibits pronounced oscillations with relatively low entanglement entropy, constrained by the symmetries present at $\alpha=0$ that are not fully broken. The entanglement entropy  exhibits a quicker and larger saturation values for intermediate $\alpha$, i.e, at $\alpha=1.0$ and $\alpha=2.0$, when compared to small $\alpha$. With further increase in $\alpha$, this entropy saturation value decreases, as observed for $\alpha=3,~4$ and $\alpha=5$. We observe a prethermal phase for large $\alpha$ ($\alpha>3$), the extent of which appears to increase with $\alpha$, reaching final steady state at larger $n$. As also seen in Fig.~\ref{entropy}(a), $\alpha=4$ reaches the final steady state within the time scales studied in this work, which is not the case for $\alpha=5$.
The time averaged entropy ($\overline{S_{N/2}}$) \textcolor{black}{normalized by $S_{\rm_{Page}}$ as a function of $\alpha$} is shown in Fig.~\ref{entropy}(b). We have not included $\alpha>4$ in the figure since the system does not reach the steady state for these large $\alpha$ values within the time scales studied. 
The above results confirm that the maximum value of $S_{N/2}$ reached in the intermediate $\alpha$ regime is close to $S_{\rm{Page}}$, pointing towards chaotic dynamics in full Hilbert space and thermalization in this regime. 
In addition, it has been shown that in systems which thermalize, the von Neumann entropy grows ballistically and saturates to its maximum value in $\mathcal O(N)$ time \cite{Khemani_Floquet}, as also seen in the intermediate range of $\alpha=1$ and $\alpha=2$ in Fig.~\ref{entropy}(a), further confirming thermalization in the intermediate $\alpha$ regime. The variation of \textcolor{black}{$\overline{S_{N/2}}/S_{\rm_{Page}}$} with  $\alpha$ for different  $\tau$ is shown in the inset of Fig.~\ref{entropy}(b). We observe a similar trend as in the $\tau=1.0$ case, where a maxima occurs at around $\alpha=2$ and the region of maxima broadens as $\tau$ is increased, consistent with $\overline{\langle J^2 \rangle}$ results. It is not clear from this figure where will the steady state value for very large $\alpha$ be with respect to $\alpha=0$ and intermediate $\alpha$, since it is difficult to obtain steady state values in very large $\alpha$ limit for the reasons mentioned above. We discuss this large $\alpha$ behavior in the next section and try to answer this question.

\subsection{For large $\alpha$}
In the previous section, we find that the total angular momentum shows large oscillations for $\alpha$ greater than 4, whereas the von Neumann entropy does not seem to show any saturation within the time range studied in this work. 
We now consider the large $\alpha$ regime corresponding to the integrable kicked Ising model, where some analytical calculations can be done for periodic boundary conditions. Although we have studied open boundary conditions in this work, it is to be noted that the boundary conditions do not matter in the thermodynamic limit, which we shall focus for comparison. These calculations will throw some light on the comparison between saturation values for different $\alpha$ regimes. Using Jordan Wigner (JW) transformation, the system can be mapped to a free fermion model and the explicit analytical calculations of $\langle J^2 \rangle$ and $S_{N/2}$ can be obtained by building upon the correlation matrices \cite{banaras_kicked_ising, diptiman_kicked_ising}. 
We shall compare the saturation values of $\overline{S_{N/2}}$ in the large $N$ limit obtained using the JW fermions, and that of analytical calculations in the permutation symmetric subspace and Page value corresponding to the full Hilbert space, and complete Fig. \ref{entropy}(b) for large $\alpha$.  We choose a simple initial spin coherent state given by $\ket{\psi_0}=\ket{0}^{\otimes N}$for our analysis. 
The Hamiltonian of the kicked Ising model is given by:
\begin{eqnarray}
H=\frac{k}{2N(\alpha)}\sum_{i}\sigma_i^x\sigma_{i+1}^x+\frac{p}{2}\sum_{n=-\infty }^{\infty }\sum_{i=1}^{N}\sigma_{i}^{z}\delta \left (\frac{t}{\tau}-n \right),
\end{eqnarray}
where $N(\alpha)=\frac{2N}{N-1}$, $p=4\pi/11$, $\tau=1.0$ and $k=6$. Let $J=\frac{k}{2N(\alpha)}$, $B=\frac{p}{2}$.
\textcolor{black}{The Jordan Wigner fermions $c_i$ is defined as:}
\begin{eqnarray}
\sigma_i^z&=&1-2c_i^{\dagger}c_i ,\nonumber \\
\sigma_i^x&=&-(c_i+c_i^{\dagger})\prod_{j<i}(1-2c_j^{\dagger}c_j).
\end{eqnarray}
The Hamiltonian in momentum space can be written as:
\begingroup
\color{black}
\begin{eqnarray}
H(t)&=&\sum_q \Psi_q^{\dagger} H_q(t) \Psi_q,\\
{\rm{with}}~H_q(t)&=&(B(t)-J\cos q)\sigma^z+J\sin q \sigma^x,
\end{eqnarray}
\endgroup
where $\Psi_q^{\dagger}=(c_q^{\dagger},c_{-q})$, $B(t)=B\sum_{n=-\infty}^{\infty}\delta(\frac{t}{\tau}-n)$ and $q=\pm \frac{1}{2}\frac{2\pi}{N} \cdots \pm \frac{N-1}{2}\frac{2\pi}{N}$ are the quasi-momenta. The two dimensional square matrices are spanned by 
$\ket{0,0}$ and $\ket{q,-q}$ corresponding to presence or absence of $c_q$ fermions \cite{banaras_kicked_ising}. 
Clearly, various modes ``$q$" in the above model are decoupled, so that each mode can be evolved independently using the Floquet evolution operator given by:
\begin{eqnarray}
U_q(\tau)&=&\exp(-iB\tau)\exp(-i\tau[(-J\cos q)\sigma^z+(J\sin q) \sigma^x]),\nonumber\\
U_q(\tau)&=&\left(
\begin{matrix}
a & -b^* \\
b & a^*
\end{matrix}
\right),
\end{eqnarray}
where $a=e^{-i2B\tau}(\cos(2J\tau))+i\cos(q)\sin(2J\tau))$ and \\$b=-i\sin(q)\sin(2J\tau)$. 
The initial state is equivalent to $\ket{\psi_0}=\prod_{q}(0,1)^T$ and
the evolved state can be written as:
\begin{eqnarray}
\begin{pmatrix}
u_q(n) \\
v_q(n)
\end{pmatrix}=
&\left(\begin{matrix}
a & -b^*  \\
b & a^*
\end{matrix}
\right)^n\begin{pmatrix}
u_q(0) \\
v_q(0)
\end{pmatrix},
\end{eqnarray}
so that the total state at time $n$ is $\ket{\psi_n}=\prod_{q}(u_q(n),v_q(n))^{T}$. The expression for $u_q$ and $v_q$ at time step $n$ is
\begin{eqnarray}
u_q(n)=-i\sin q \sin(2J\tau)\frac{\sin n\theta_q}{\sin \theta_q}, \nonumber \\
v_q(n)=\cos n\theta_q+i \mathrm {Im}(a^*)\frac{\sin n\theta_q}{\sin \theta_q},
\end{eqnarray}
where, \[\theta_q=\cos^{-1}(\cos(2B\tau) \cos(2J\tau)-\cos q \sin(2B\tau) \sin (2J\tau)).\]
The fermionic correlators are given by \cite{banaras_kicked_ising}:
\begin{eqnarray}
C_{ij}&=&\langle c_i^\dagger c_j \rangle=\frac{2}{L}\sum_{q>0}|u_q(n)|^2 \cos q(i-j) ,\nonumber \\
F_{ij}&=&\langle c_i^\dagger c_j^\dagger \rangle= \frac{2}{L}\sum_{q>0}u_q^*(n)v_q(n) \sin q(i-j).
\label{eq:correlators}
\end{eqnarray}
We first look at the calculations of $\langle J^2 \rangle$. An analysis based on correlation functions can be used to calculate $\langle J^2 \rangle$, as discussed below. In terms of Pauli matrices, $\langle J^2 \rangle$ can be written as:
\begin{eqnarray}
\langle J^2 \rangle =\left \langle \frac{1}{2}\sum_{i < j} (\sigma_i^x\sigma_j^x+ \sigma_i^y\sigma_j^y+\sigma_i^z\sigma_j^z)+  \frac{3N}{4}\mathbb{I}_{2^N}\right \rangle
\end{eqnarray}
Thus, to evaluate $\langle J^2 \rangle$ we need the spin spin correlation functions $\langle \sigma_i^x \sigma_j^x\rangle$, $\langle \sigma_i^y \sigma_j^y\rangle$ and $\langle \sigma_i^z \sigma_j^z\rangle$.
Since $\langle \sigma_i^z \sigma_j^z\rangle$ does not involve the Jordan Wigner non local string, it can be estimated easily as: 
\begin{eqnarray}
\langle \sigma_i^z \sigma_j^z \rangle &=& \langle (1-2c_i^\dagger c_i)(1-2c_j^\dagger c_j)\rangle, \nonumber \\
&=& \langle 1-2c_ic_i^\dagger-2c_j c_j^\dagger +4 c_i^\dagger c_i c_j^\dagger c_j             \rangle \nonumber, \\
&=&1-4\langle c_i^\dagger c_i \rangle+4\langle c_i^\dagger c_i\rangle^2+4|\langle c_i^\dagger c_j^\dagger \rangle|^2-4|\langle c_i^\dagger c_j \rangle|^2 \nonumber. \\
\end{eqnarray}
where the $\langle c_i^\dagger c_j \rangle$ and $\langle c_i^\dagger c_j^\dagger \rangle$ are given in Eq.~\ref{eq:correlators}.
We now evaluate the other set of correlation functions:
\begin{eqnarray}
\langle \sigma_i^x \sigma_j^x \rangle &= -(c_i+c_i^{\dagger})\prod_{k<i}(1-2c_k^{\dagger}c_k)(-(c_j+c_j^{\dagger}) \nonumber \\
&\qquad \prod_{k'<j}(1-2c_{k'}^{\dagger}c_{k'})), \nonumber \\
&=\langle(c_i+c_i^\dagger)\prod_{k=i}^{j-1}(1-2c_k^\dagger c_k)(c_j+c_j^\dagger)\rangle.
\end{eqnarray}
Let $A_j=c_j^\dagger +c_j$ and $B_j=c_j^\dagger-c_j$ giving $A_jB_j=1-2c_j^\dagger c_j$.

\begin{figure}[ht]
\centering
    \includegraphics[width=1.0\linewidth, height=0.8\linewidth]{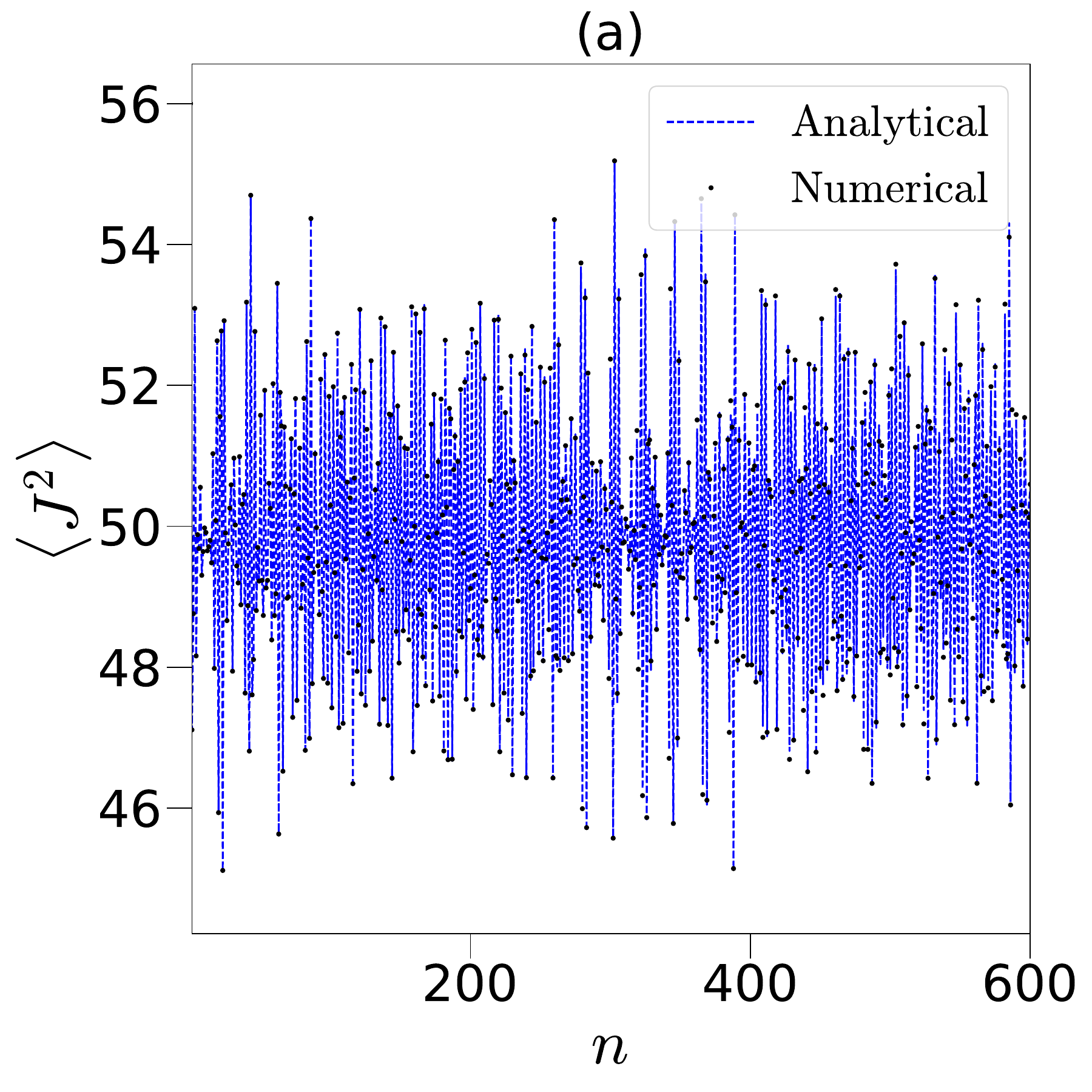}\\
     \includegraphics[width=1.0\linewidth, height=0.8\linewidth]{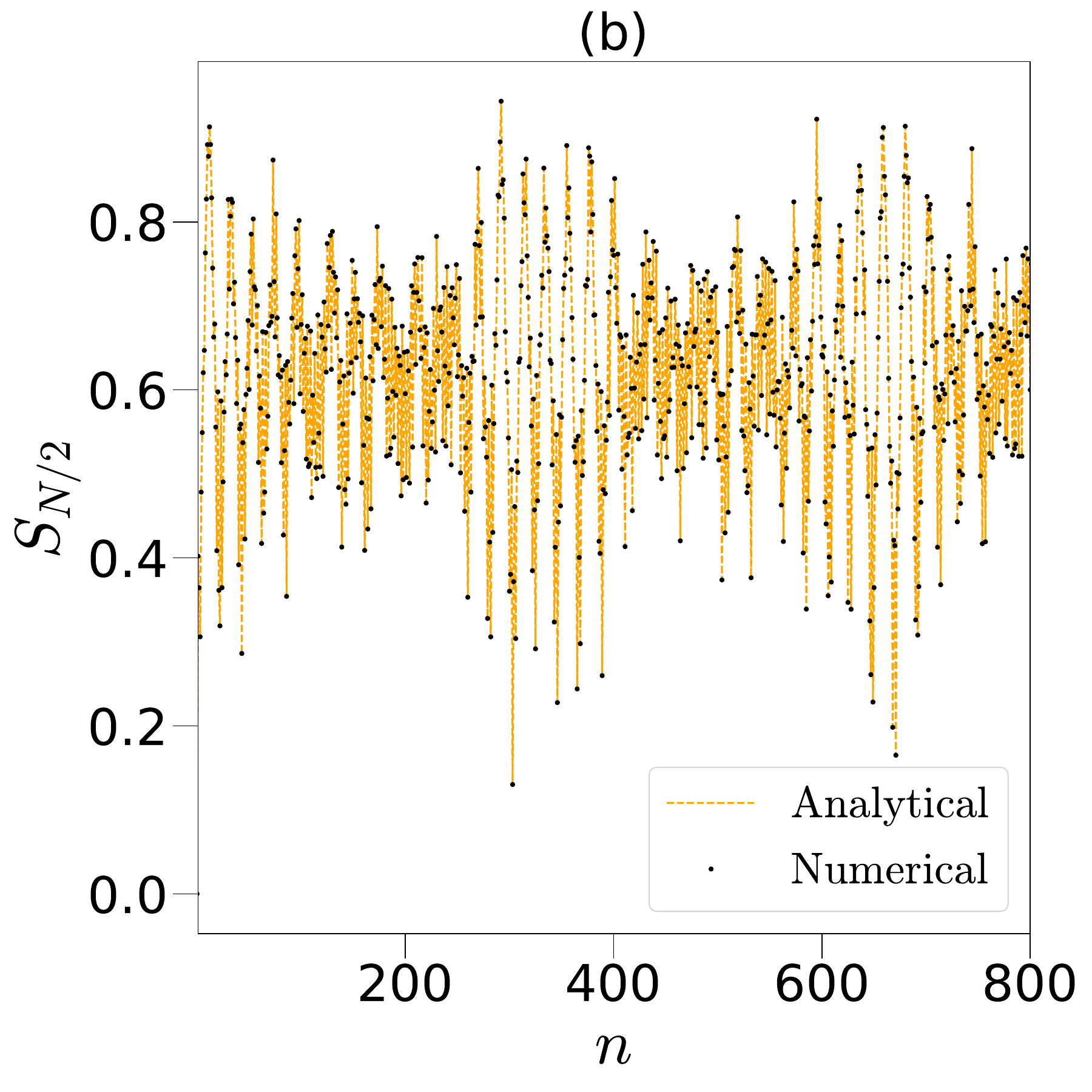}
   \caption{(a). The plot of $\left \langle J^2 \right \rangle$ with respect to time $n$ for $N=14$ at $\tau=1.0$ for $\alpha=15$. The dashed line represents the analytical results and dots correspond to the numerical results. (b). The plot of $S_{N/2}$ with respect to $n$ for $N=14$ at $\tau=1.0$ and $\alpha=15$. The dashed line represents the analytical results and the dots represents the numerical values. The initial state is $\ket{0}^{\otimes N}$. }

  \label{kicked_ising}
\end{figure}

Thus,
\begin{eqnarray}
\langle \sigma_i^x \sigma_j^x \rangle &=&\langle A_i\prod_{k=i}^{j-1}A_kB_kA_j   \rangle, \nonumber \\
&=& \langle B_i A_{i+1} B_{i+1}.......A_{j-1}B_{j-1}A_j \rangle.
\end{eqnarray}
Using Wicks theorem, the expectation can be expressed in the form of Pfaffian of a matrix as \cite{scipost, spin_spin_correlation}:
\begin{eqnarray}
\langle \sigma_i^x \sigma_j^x \rangle= (-1)^{r(r+1)/2}Pf(M),
\end{eqnarray}
where $r=j-i$ and
\begin{eqnarray}
M=\left(
\begin{matrix}
X_{r \times r} & Z_{r \times r} \nonumber \\
-Z^T_{r \times r} & Y_{r \times r}
\end{matrix}
\right) .\\
\end{eqnarray}
For $m\neq n$,
\begin{eqnarray}
X_{mn}&=&\langle A_mA_n \rangle, \nonumber\\
Z_{mn}&=&\langle A_mB_{n-1} \rangle, \nonumber\\
Y_{mn}&=&\langle B_mB_n \rangle. \nonumber 
\end{eqnarray}
which are given by:
\begin{eqnarray}
\langle A_mA_n \rangle &=& 2i  {\mathrm{Im}}(\langle c_m c_n \rangle+\langle c_m^{\dagger}c_n \rangle),\nonumber\\
\langle A_mB_{n-1} \rangle&=& 2 \mathrm{Re}(\langle c_m^{\dagger}c_{n-1} \rangle-\langle c_m c_{n-1}\rangle)-\delta_{mn-1}, \nonumber\\
\langle B_mB_n \rangle&=& 2i \mathrm{Im}(\langle c_m c_n \rangle-\langle c_m^{\dagger}c_n \rangle). \nonumber 
\end{eqnarray}
Similarly we can also find the spin spin correlations $\langle \sigma_i^y \sigma_j^y \rangle$ along y-direction.
The plot of $\langle J^2 \rangle$, evaluated both numerically and from the semi analytical calculations described above is shown in Fig.~\ref{kicked_ising}(a). It shows a close match between the analytical and the numerical results.

\begin{figure}[ht]
\centering
\includegraphics[width=0.95\linewidth, height=0.8\linewidth]{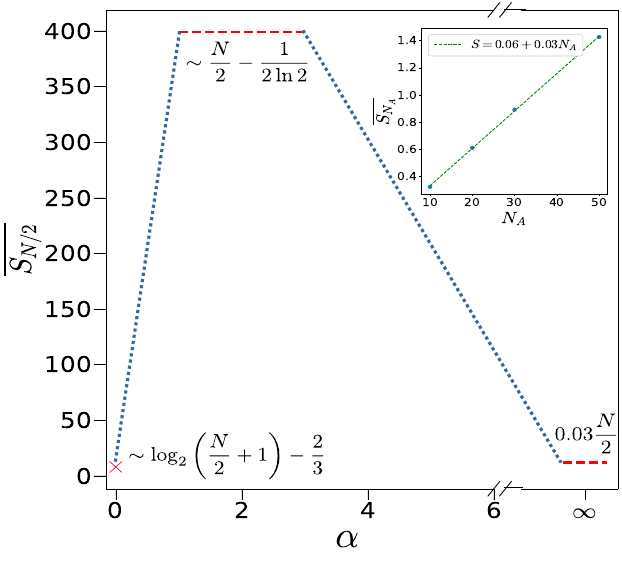}
\caption{\textcolor{black}{The plot shows the schematic of $\overline{S_{N/2}}$ with $N=800$ for $\alpha=0$ corresponding to random permutation symmetric subspace in $N+1$ dimensions and the intermediate range governed by the Page value in $2^N-$dimensional Hilbert space. For $\alpha \to \infty$ corresponding to the kicked Ising model, $\overline{S_{N_{A}}}$ for different subsystem size $N_A$ is computed using semi-analytics for $\tau=1$ as shown in the inset, which is then used to obtain $\overline{S_{N/2}}$ from the fitting.}}
\label{relative_strength}
\end{figure}

The von Neumann entropy $S_{N/2}$ can also be evaluated using the eigenvalues of correlator matrix given by:
\begin{eqnarray}
\Pi_n(l)=\left(\begin{matrix}
I_{l}-C & F  \\
F^* & C
\end{matrix}
\right),
\end{eqnarray}
where $l$ is the subsystem size \cite{banaras_kicked_ising}. Here, $C$ and $F$ matrices are restricted to within the subsystem size and are defined above in Eq.~\ref{eq:correlators}.
The agreement between the analytical and the numerical results is shown in Fig.~\ref{kicked_ising}(b). We also observe that the amplitude of oscillations decreases as the system size increases. Therefore, we use the large $N$ behavior where the boundary conditions are irrelevant, to sketch $\overline {S_{N/2}}$ vs $\alpha$, comparing analytical expressions corresponding to permutation symmetric subspace ($\alpha=0$), Page value (intermediate $\alpha$), and the value obtained using semi analytical approach for large $\alpha$, see Fig. \ref{relative_strength}. For large $\alpha$, previous studies have shown that the saturation values of entropy scales linearly with subsystem size \cite{russomanno2016entanglement}, consistent with our results presented in the inset of Fig.~\ref{relative_strength}. We have used this linear fitting to extrapolate the value of $S_{N/2}$ for $N=800$, which otherwise is numerically time consuming. It clearly shows the steady state values relative to each other for the three different range of $\alpha$. A similar trend should be seen for finite system sizes as well.

\section{Floquet eigenstates}
\label{floquet_eigenstates}
In the previous sections, we studied the behavior of dynamical quantities like $\langle J^2 \rangle$ and $S_{N/2}$ with respect to $\alpha$. We have seen that for $\alpha$ close to zero, the long time averaged values of these quantities closely match the  results predicted by RMT within the permutation symmetric subspace of $N+1$ dimensions. In the intermediate $\alpha$ regime, these quantities thermalize and saturate to values corresponding to RMT in full Hilbert space of dimension $2^N$. On the other hand, the large $\alpha$ limit corresponding to kicked Ising model does not thermalize due its integrability. We have also seen that for large $\tau$, the system tend to thermalize at smaller $\alpha$, and continues to exhibit this thermal behavior for a wider interval of $\alpha$ . We now look into the eigenstate properties of the system in order to identify the regimes where thermalization occurs, and explain the numerically obtained results discussed above. We mainly use two quantities, the effective dimension $D_{\rm eff}$ and the spectral statistics, to better understand the dynamics of the system as a function of $\alpha$.

\subsection{Effective Dimension $\rm D_{eff}$}
\begin{figure}[ht]
\centering
   \includegraphics[width=1.0\linewidth, height=0.8\linewidth]{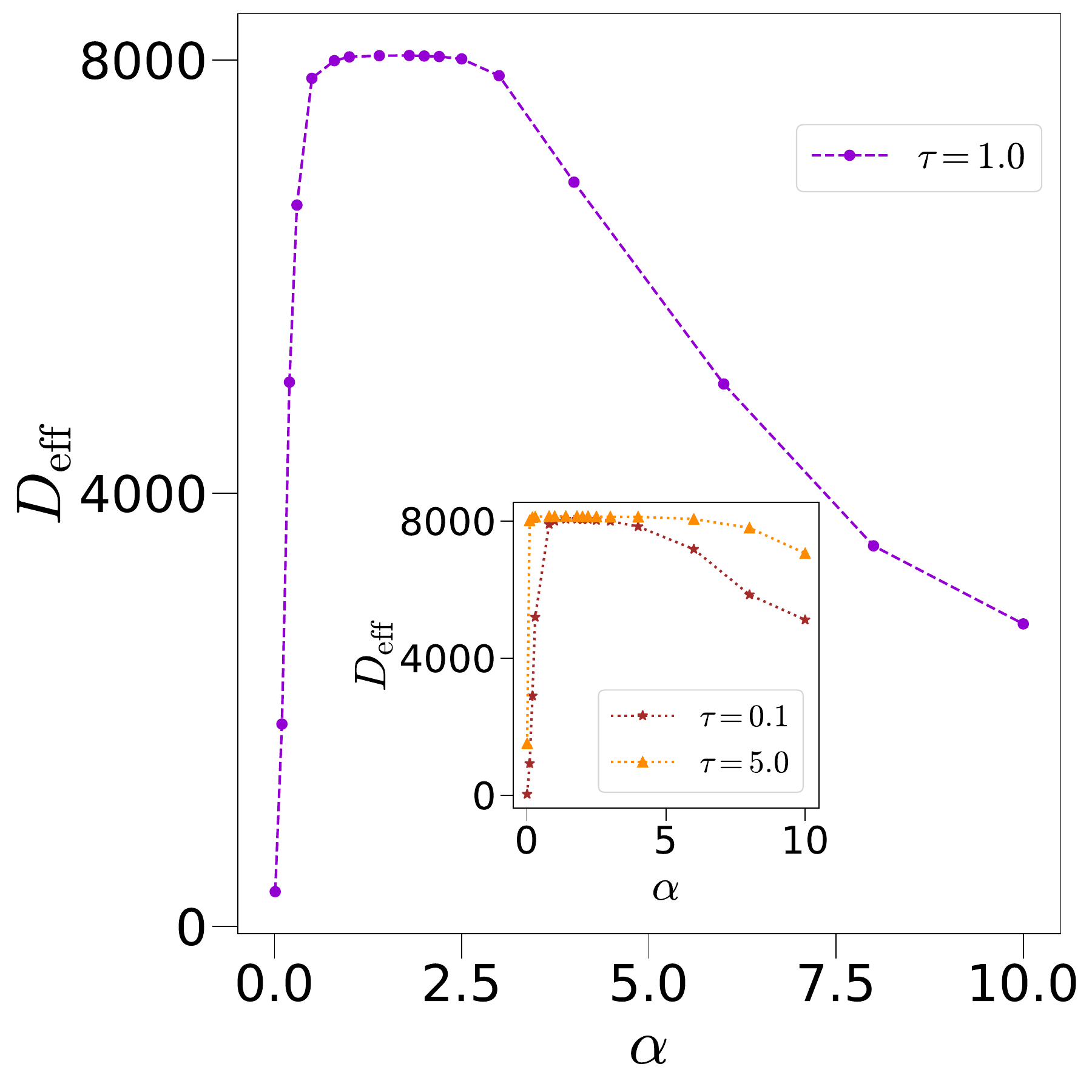}
   \caption{\textcolor{black}{The plot of $ D_{\rm eff} $ with respect to $\alpha$ for $N=14$ at $\tau=1.0$. The inset shows $D_{\rm eff}$ for different $\tau$ values at same parameter values. }}

  \label{D_eff}
\end{figure}

We first look at the quantity called effective dimension $D_{\rm eff}$ \cite{manju25}. To calculate this, we express the initial spin coherent state $\ket{\theta,\phi}$ in terms of Floquet eigen states $\ket{\phi_{i}}$ obtained from the diagonalization of the Floquet operator $U$ (Eq. \ref{eq:Floquet}), as shown below:
\begin{eqnarray}
\ket{\theta \phi }= \sum_{i=1}^{2^N} c_{i}\ket{\phi_{i}}.
\end{eqnarray}
We obtain $c_{i}$ by computing the overlap of the initial state with the Floquet eigenstates $|\phi_{i}\rangle$. The effective dimension $D_{\rm eff}$ is defined as that value of $K$ for which  $\sum_{i=1}^{K}|c_{i}|^{2}=1-\epsilon$ where $\epsilon$ is a small parameter,  after arranging the coefficients $c_{i}'s$ in decreasing order of their magnitude \cite{Haakebook}. The plot of $D_{\rm eff}$ with respect to $\alpha$ is shown in Fig.~\ref{D_eff} for $\tau=1$ and $\epsilon=0.0001$. We see that $D_{\rm eff}\sim N+1$ for $\alpha=0$ corresponding to the dimension of permutation symmetric subspace. As $\alpha$ is increased, $D_{\rm eff}$ increases and reaches close to $2^{N-1}$ for intermediate range of $\alpha$ (between $1$ and $2.5$), and then decreases when $\alpha$ is increased further. This decrease in $D_{\rm eff}$ for large $\alpha$ is expected since the system approaches the integrable kicked Ising model in this limit, where due to the presence of conserved quantities the system may not explore the entire accessible Hilbert space. The dependence of the maximum value of $D_{\rm eff}$ in the intermediate regime on the parameter $\epsilon$ can be estimated using random states as presented in our earlier work \cite{manju25}. The only difference being now the total Hilbert space dimension is less than $2^N$ due to the bit-reversal symmetry as discussed below.
It stems from the invariance of the system under swapping of spins at site $i$ and $N-i+1$ for  $i=1,2, \cdots N$ \cite{Rohit_sunil_mishra_ising}. The eigenvalues of the bit reversal operator $\hat{B}$ are $\pm1$. Thus, the eigenstates can be classified as even or odd under bit reversal. The initial spin coherent state having an even parity under bit reversal gives non zero overlap with only the eigenstates within the same parity sector. The dimension of the even parity sector $B_{+}=2^{N-1}+2^{N/2-1}$ \cite{bit_reversal_2024}, is close to the $D_{\rm eff}$ obtained numerically in the intermediate $\alpha$ regime. The variation of $D_{\rm eff}$ with respect to $\alpha$ for different values of $\tau$ is shown in the inset of Fig.~\ref{D_eff}. It is consistent with the behavior of other dynamical quantities, i.e., with increase in $\tau$, the $D_{\rm eff}$ attains the maximum value close to $B_+$ dimension for smaller $\alpha$ and remains close to this over a broader range of $\alpha$.

It is to be noted that for the evaluation of $J^2$ for random pure state in full Hilbert space dimension, the $D_{\rm eff} \approx 2^N$. However, for our model, even though $D_{\rm eff} \approx 2^{N-1}$, the total angular momentum $\langle J^2 \rangle$ in the intermediate $\alpha$ regime still approaches to the full Hilbert space RMT value $3N/4$. The details are discussed in Appendix \ref{app_bitreversal}. 

\subsection{Spectral Statistics}
\label{sec_spectralstatistics}
\begin{figure}[ht]
\centering
 \includegraphics[width=1.0\linewidth, height=0.8\linewidth]{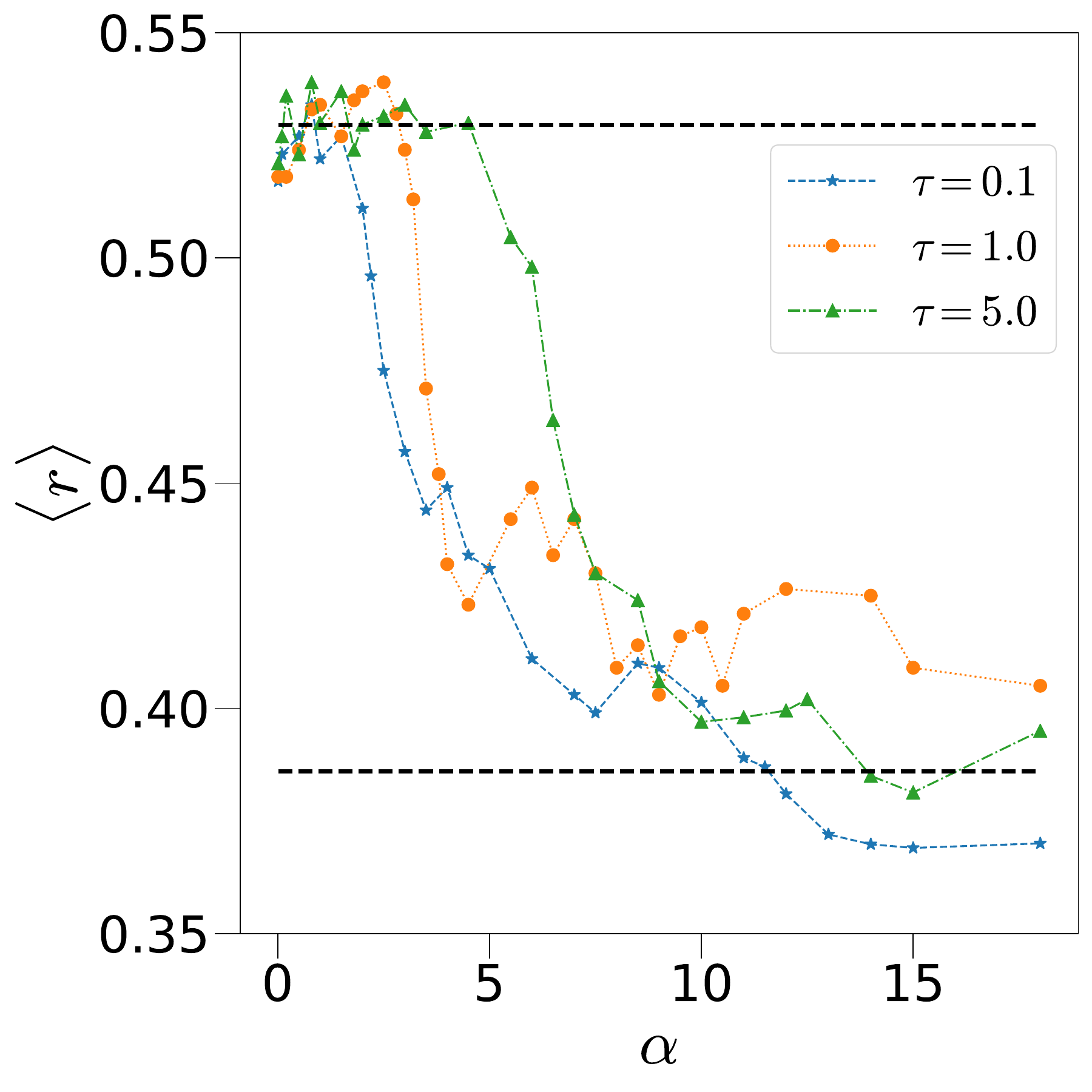}
   \caption{\textcolor{black}{The average level spacing ratio $\langle r \rangle$ with respect to $\alpha$ for $N=14$ at different values of $\tau$. The black dashed lines represents the average values $\langle r \rangle_{WD}=0.529$ and $\langle r \rangle_{P}=0.386$ corresponding to the Wigner Dyson (COE) and Poisson distributions respectively.}}

  \label{statistics_ratio}
\end{figure}

We now look at the behavior of Floquet spectrum using spectral level statistics at different values of $\alpha$. This can be used to differentiate between chaotic and integrable regime as we will see in this section. We consider the level spacing ratios $r_n$ defined by \cite{statistics_oganesyan, PRL_statistics, wang2023statistics}:
\begin{equation}
r_n=\frac{\min \{s_n,s_{n-1} \}}{\max \{s_n,s_{n-1} \}}.
\end{equation}
Here, $s_n=E_{n+1}-E_{n}$ denotes the differences between consecutive energy levels, with $E_n$ being the $n$th eigenphase of the Floquet operator $U$ given in Eq.~\ref{eq:Floquet}. Since, the Floquet operator $U$ commutes with parity operator $\hat{R}=e^{-i\pi J_z}$, the matrix can be block diagonalized into even and odd parity blocks. In addition, the model also has bit reversal symmetry, resulting in eigenstates corresponding to odd or even under bit reversal \cite{Rohit_sunil_mishra_ising, bit_reversal_2024}. For the analysis, we use eigenstates corresponding to even parity under both $\hat{R}$ and bit reversal. The mean level spacing ratio, defined as $\langle r \rangle=\frac{1}{N}\sum_{i}^{N}r_{i}$, distinguishes between integrable and chaotic systems. For the integral systems showing Poisson statistics, $\langle r \rangle_{P}=0.386$ and for the chaotic system with Wigner Dyson distribution corresponding to the circular orthogonal ensemble (COE), $\langle r \rangle_{WD}=0.529$ \cite{wang2023statistics, ensemble_inequivalence}.
We plot $\langle r \rangle$ as a function of $\alpha$ in Fig. \ref{statistics_ratio}.
It is well established that for $\alpha=0$, the kicked top model at large values of $k$ exhibits chaotic behavior and the nearest neighbor statistics follows the Wigner Dyson distribution corresponding to the COE ensemble \cite{wang2023statistics}. We see that this trend continues as we increase $\alpha$ up to around $2$ for $\tau=1$, beyond which the spectrum deviates away from the Wigner Dyson behavior and approaches the Poisson statistics for large values of $\alpha$ showcasing the integrable behavior of kicked Ising model as $\alpha \rightarrow \infty$. We repeat the same calculations for other values of $\tau$. We find that the system continues to show Wigner Dyson statistics till larger values of $\alpha$ for large $\tau$, after which the statistics switches to Poisson. This is also consistent with the fact that the thermal phase extends upto large $\alpha$ for large $\tau$ values as the system is chaotic upto larger $\alpha$ for these $\tau$ values. 
\section{System-size dependence}
\label{sec_N}
\begin{figure}[ht]
\centering
\includegraphics[width=1.0\linewidth, height=0.8\linewidth]{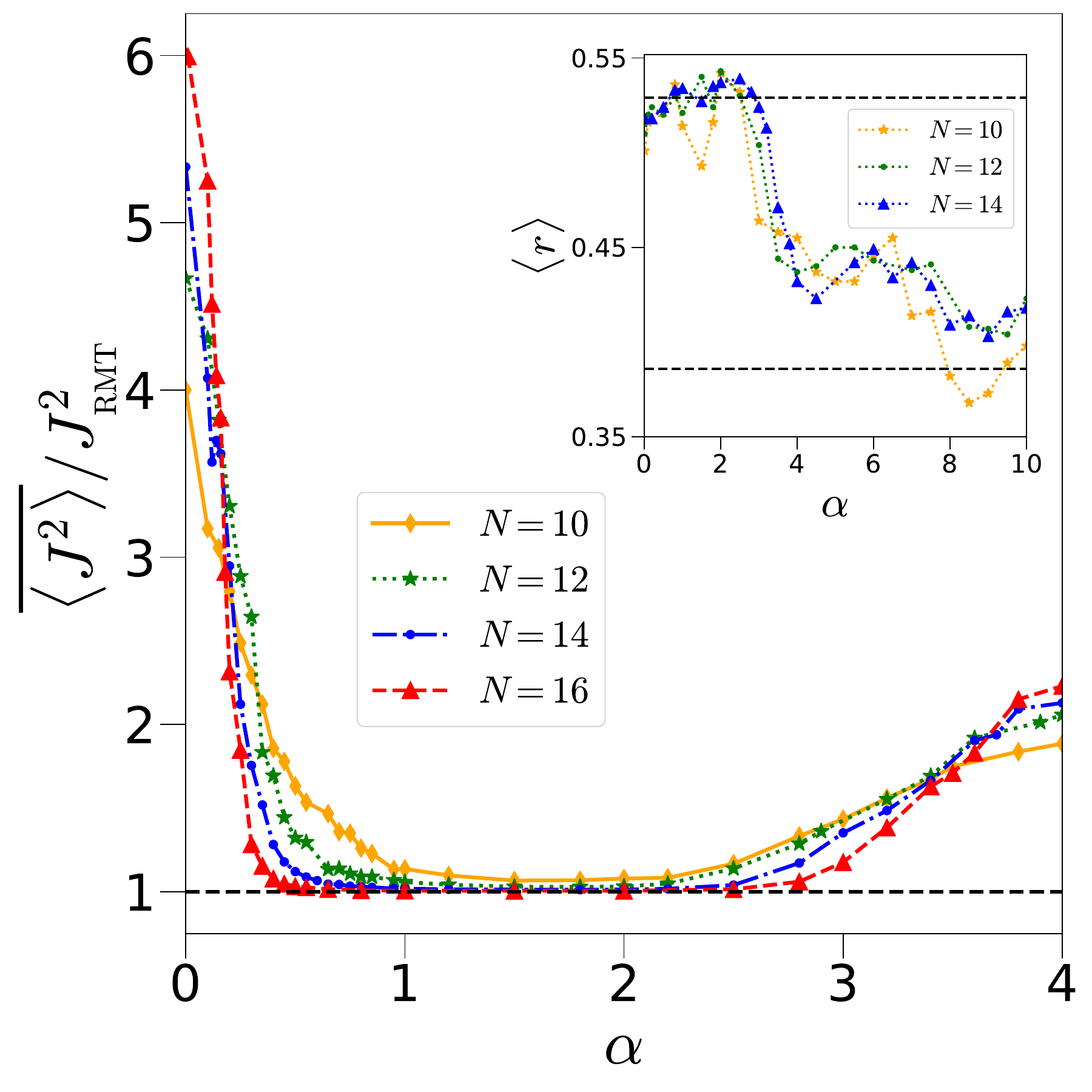}
   \caption{\textcolor{black}{The plot shows the time averaged $\langle J^2 \rangle$ scaled by $J^2_{\rm_{RMT}}$ as a function of $\alpha$ at $\tau=1$ for different system sizes $N$. The inset presents a similar analysis for the average spacing ratio $\langle r \rangle$ plotted against $\alpha$ for different $N$ at $\tau=1$. Together, these results clearly illustrate the trend towards thermalization as a function of $\alpha$ with increase in $N$.}}

  \label{diff_system_size}
\end{figure}

\textcolor{black}
{Till now, we have analyzed dynamical observables and Floquet eigen-state properties for a fixed system size $N=14$ and explored its behavior as a function of the driving period $\tau$. We now investigate how thermalization behavior changes with varying system sizes ($N=10,~12,~14,~16$). In Fig.~\ref{diff_system_size}, we plot the time averaged $\overline{\langle J^2 \rangle}$ with respect to $\alpha$ for different system sizes $N$. We observe that as the system size $N$ increases, the range of intermediate $\alpha$ values for which the system reaches thermalization becomes broader, indicating that larger systems thermalize over an increasingly wide interval of intermediate $\alpha$.
 A further confirmation of this trend is shown in the inset of Fig.~\ref{diff_system_size} where $\langle r \rangle$ is plotted with respect to $\alpha$ for different system sizes. We see that the largest system size considered, i.e., $N=14$ shows 
 $\langle r \rangle_{WD}$ upto larger  $\alpha$ as compared to smaller system sizes, and the approach to the Poisson value $\langle r \rangle_P$ of the integrable limit occurs at larger $\alpha$ values for larger $N$. It is to be noted that $\langle r \rangle$ shows a value smaller than $\langle r \rangle_P$ for $N=10$, which is not the case for $N=12$ and $14$, pointing towards possible finite size effects, as also pointed out in similar previous studies \cite{ensemble_inequivalence}.
 The results from the analysis of $J^2$ and $\langle r \rangle$ suggest that for sufficiently large $N$, the system will thermalize for any finite, non-zero $\alpha$. This behavior aligns with the expectation that a generic interacting Floquet system, in the absence of integrability, will absorb energy from the drive and eventually reach thermalization in the long-time limit. \cite{karthik_thermalization_floquet,PRE_RMT_infinite_temp, exponential_slow_heating}}.

\section{Conclusion}
\label{conclusion}

In this work, we study a long range interacting spin system where the strength of interaction between any two spins at a distance $r$ decreases as $1/r^{\alpha}$, with $\alpha$ tuning the range of interactions. For $\alpha=0$, it is an all-to-all interacting system having permutation symmetry. We focus on breaking this symmetry by switching on a non-zero $\alpha$, and study its effect on various dynamical quantities like the total angular momentum $J^2$ and von Neumann entropy $S_{N/2}$. Fixing the chaos parameter at $k=6$ corresponding to the chaotic limit of permutationally symmetric $\alpha=0$ model, as $\alpha$ is gradually turned on \textcolor{black}{for a fixed system size $N$}, the system exhibits near permutation symmetric behavior for small $\alpha$ where the steady state values are given by the random state results in symmetric subspace of $N+1$ dimensions. For very large $\alpha$ on the other hand, it shows integrable kicked Ising dynamics. 
 The study also shows the existence of thermalizing regime in the intermediate $\alpha$ range where the steady state values of $J^2$ and $S_{N/2}$ closely match with the random state results in full $2^N$ dimension given by the random matrix theory.
We also examine the dependence of thermalization on different driving periods $\tau$. We find that for large $\tau$, thermalization \textcolor{black} {to random state values corresponding to  full Hilbert space} sets in at lower $\alpha$, and persists for a larger range of $\alpha$ due to easier absorption of energy at large $\tau$. The study using Floquet eigenstates show that for small $\alpha$, $D_{\rm eff}\sim N+1$ and for the intermediate regime where the steady state values show extremum as a function of $\alpha$, the effective dimension exhibits maximum value close to $2^{N-1}$, signaling extensive delocalization of Floquet eigenstates in this region. For large $\alpha$, $D_{\rm eff}$ decreases indicating the presence of conserved quantities for the integrable kicked Ising model. Finally, the spectral statistics also show features, qualitatively similar to other quantities studied, with the average spectral ratio $\langle r \rangle$ showing a transition from the COE ensemble of the random matrix theory to the Poisson value of the integrable limit as $\alpha$ is tuned from $0$ to large values. The fact that the average spectral ratio $\langle r \rangle$ converges to the Wigner Dyson value in the intermediate $\alpha$ regime confirms the chaotic and hence the thermalizing behavior in this region. \textcolor{black}{We also studied the system-size dependence of thermalization to random state corresponding to full Hilbert space. We find that as the system size increases, the range of $\alpha$ for which this thermalization occurs also increases, indicating a possibility of thermalization for any non-zero $\alpha$ in the limit of very large $N$.}

Although, we have performed our analysis with $k$ set to $6$, and an initial coherent state $\ket{\theta=2.25,~\phi=1.1}$, the results remain qualitatively similar for different orientations $(\theta,\phi)$ of the initial spin coherent state. Comparable results are also obtained for different values of kicking strength $p$, where additional symmetries, if present, may need to be taken into account for the spectral analysis. We fix $k=6$ throughout our analysis, since lower $k$ corresponds to regular region of the kicked top model ($\alpha=0$), having strong oscillations that prevent the reliable determination of the steady state saturation values of dynamical quantities. Preliminary checks using $D_{\rm eff}$ suggest that for $k=1$, the system does not reach the exponentially large Hilbert space dimension ($\sim 2^{N}$), and the spectral statistics does not exhibit a Wigner Dyson distribution for any $\alpha$ in the intermediate regime. A detailed analysis of these low $k$ regimes is left for future work. To conclude, our study finds the possibility of thermalization to random state corresponding to the full Hilbert space for intermediate range of tuning parameter $\alpha$ provided the corresponding fully permutation symmetric model ($\alpha=0$) lies in the chaotic regime. This range of thermalization can be controlled by the period of kicks. Spectral statistics and effective dimension $D_{\rm eff}$ corroborate these observations.

{Compared to previous works related to long range Floquet systems that studied thermalization, prethermalization and stability of dynamical phases in related but distinct Hamiltonian models \cite{kapitza, santos_heating_suppression, exponential_slow_heating, classical_spin_floquet, pizzi_higher_order}, our work focuses on the effect of breaking permutation symmetry in an otherwise all-to-all interacting model. Other than the usual von Neumann entropy studied, we also calculate total angular momentum $J^2$ which is a constant of motion in the $\alpha=0$ limit, and thus is a good measure to understand how far the system is from the permutation symmetric state. In addition, we also examine $D_{\rm eff}$ that reveals the departure of the dynamics from the symmetric subspace as $\alpha$ is tuned. We provide a clear characterization of the transition from Wigner Dyson to Poisson as a result of this symmetry breaking. Thus, it provides a different perspective to existing studies emphasizing the role of symmetry constraints in Floquet systems.}
There are recent works \cite{puri2024floquet}, related to long range interactions along with Floquet driving leading to enhancement of the performance of quantum batteries. Thus, we believe that our study provides valuable insights to quantum technologies where the interplay of interaction strength and Floquet driving is essential to bring out the maximum output from the system. 

\appendix

\section{Bit reversal symmetry}
\label{app_bitreversal}
In this section, we show that $J^2$ attains value close to full Hilbert space of $2^N$ dimension for random state dynamics even in the presence of bit reversal symmetry. As discussed in Ref. \cite{Rohit_sunil_mishra_ising}, the dimension of even parity sector is given by $2^{N-1}+2^{N/2-1}$, and that of odd parity sector is $2^{N-1}-2^{N/2-1}$.  For our analysis, we consider only the even parity sector of the bit reversal as the initial spin coherent state is even under bit reversal.

If $\ket{\Phi_\ell}=\ket{j_1 \cdots j_N}$, $j_i=\pm 1$,
are the eigenvectors of $J_z$ with eigenvalues $\lambda_\ell=(\sum_{i} j_{i})/2$, the expectation value of $J_z^2$ in a random state $\ket{\Psi}_{\rm RMT}$ is given by:
\begin{equation}
\begin{split}
\left \langle\ J_{z}^2 \right \rangle_{{\text{RMT}}} &=\sum_{\ell=1}^{2^N}\lambda_{\ell}^2 \overline{|\left \langle \Psi_{\text{RMT}}| \Phi_{\ell} \right \rangle|^{2}}, \\
&=\sum_{l=1}^{B_{+}}\lambda_l^2 \left(\frac{1}{2^{N-1}+2^{N/2 -1}} \right).
\label{eq_jzrmt}
\end{split}
\end{equation}
The ensemble average over random states $\overline{|\left \langle \Psi_{\text{RMT}}| \Phi_{\ell} \right \rangle|^{2}}=1/(2^{N-1}+2^{N/2 -1})$, due to the constraint imposed by the bit reversal symmetry. Evaluating the sum $\sum_{l=1}^{B_{+}} \lambda_l^2$ is equivalent to computing the $\rm tr$$(J_z^2)$ in the even parity sector of the bit reversal symmetry, and is given by \cite{bit_reversal_2024}:
\begin{equation}
\sum_{l=1}^{B_{+}}\lambda_l^2=\sum_{B_{+}} \mathrm {tr}(J_z^2)=2^NN/8+ 2^{N/2}N/4 .
\end{equation}
Substituting in Eq. \ref{eq_jzrmt}, we get
\begin{align}
\langle\ J_{z}^2 \rangle_{{\text{RMT}}} &=\frac{1}{(2^{N-1}+2^{N/2 -1})} (2^NN/8+2^{N/2}N/4), \nonumber \\
&=\frac{N}{4}\frac{\left(1+\frac{2}{2^{N/2}}\right)}{\left(1+\frac{1}{2^{N/2}}\right)}, \nonumber \\
&\approx N/4.
\end{align}
Similarly  $\left \langle J_x^2 \right \rangle_{\text{RMT}}=\left \langle J_y^2 \right \rangle_{\text{RMT}}\approx N/4$, and hence $\left \langle J^2 \right \rangle_{\text{RMT}}\approx3N/4$. \textcolor{black}{For convenience, we denote $\langle J^2 \rangle_{\rm_{RMT}}$ by $J^2_{\rm_{RMT}}$ in the main text.} 

\begin{acknowledgments}
    MC and UD acknowledge Arul Lakshminarayan for collaboration in related works. MC acknowledges Lea F. Santos for valuable discussions during ``Long-Range Interactions and Dynamics in Complex Quantum Systems”, workshop held in Nordita, Stockholm, 2025. MC also thanks Rohit Kumar Shukla for related discussions.
\end{acknowledgments}


%

\end{document}